\documentclass{article}
\usepackage{spconf,amsmath,amssymb,graphicx,url,tabularx,multirow,enumitem,hyperref}
\usepackage[dvipsnames]{xcolor}

\title{Joint Speech Recognition and Audio Captioning}
\name{
\begin{tabular}{c}
Chaitanya Narisetty$^{\star}$, Emiru Tsunoo$^{\dagger}$, Xuankai Chang$^{\star}$, Yosuke Kashiwagi$^{\dagger}$, \\ \textit{Michael Hentschel}$^{\dagger}$, \textit{Shinji Watanabe}$^{\star}$
\end{tabular}
}
\address{$^\star$ Carnegie Mellon University, USA,\\ $^\dagger$ Sony Group Corporation, Japan 
}
\begin{document}
\ninept
\maketitle
\begin{abstract}
Speech samples recorded in both indoor and outdoor environments are often contaminated with secondary audio sources.
Most end-to-end monaural speech recognition systems either remove these background sounds using speech enhancement or train noise-robust models.
For better model interpretability and holistic understanding, we aim to bring together the growing field of automated audio captioning (AAC) and the thoroughly studied automatic speech recognition (ASR).
The goal of AAC is to generate natural language descriptions of contents in audio samples.
We propose several approaches for end-to-end joint modeling of ASR and AAC tasks and demonstrate their advantages over traditional approaches, which model these tasks independently. 
A major hurdle in evaluating our proposed approach is the lack of labeled audio datasets with both speech transcriptions and audio captions. 
Therefore we also create a multi-task dataset by mixing the clean speech Wall Street Journal corpus with multiple levels of background noises chosen from the AudioCaps dataset. 
We also perform extensive experimental evaluation and show improvements of our proposed methods as compared to existing state-of-the-art ASR and AAC methods.
\end{abstract}
\section{Introduction}
\label{sec:intro}
Automatic speech recognition (ASR) is a well known task which decodes human speech as textual representations and has been a prominent research area for the past several decades \cite{yu2016automatic}.
End-to-end (E2E) ASR is a sequence-to-sequence task where a stream of acoustic features are converted into a sequence of words.
These models typically use an encoder-decoder framework with connectionist temporal classification (CTC) loss \cite{graves2014towards, amodei2016deep} or use RNN-Transducers \cite{graves2012sequence,li2020developing}.
The introduction of attention mechanisms enabled the decoder to focus on relevant parts of the audio signal when generating long word sequences \cite{chorowski2015attention,chan2016listen}.
Transformer layers with self-attention in both the encoder and decoder, have shown significant improvement in training speeds and modeling long-range dependencies \cite{karita2019comparative}.

Speech utterances captured in the real-world coexist with a wide variety of acoustic sources and are seldom pure.
However most ASR systems focus only on the speech component of an audio signal and consider the non-speech sources as components to be disregarded during model learning \cite{li2014overview,haeb2019speech}.
Synchronously modeling both speech and non-speech sources in a unified manner resembles the natural perception of the human auditory system.
Such a unified model using Transformers was proposed by \cite{moritz2020all} to solve ASR, acoustic event detection (AED) and audio tagging (AT) tasks.
Although tasks like AED and AT extract information of all constituent audio sources, the outputted sequence of events and tags lacks proper structure for straightforward human understanding \cite{kim2019audiocaps}.

Automated audio captioning (AAC) aims to provide the information of constituent audio sources and events in a structured and easily comprehensible manner i.e., a natural language description of a given audio waveform \cite{kim2019audiocaps,drossos2017automated}.
Recently, Transformer based encoder-decoder frameworks are being employed to model the temporal structure of audio events \cite{narisetty2021_t6,mei2021audio}
AAC is an emerging research area with several applications, such as enriching the raw textual information provided by ASR during television broadcasting and video streaming.
Such integration of ASR and AAC tasks can potentially improve the viewing experience of the hearing impaired. 

Performance of an ASR model in a noisy environment depends on its ability to adapt to various background sounds and robustly infer the speech components.
This process makes the ASR and AAC tasks interdependent and motivates us to treat them in a unified manner.
Also, given the similarity of these tasks in transforming input acoustic signals into output word sequences, we formulate them as a multi-task problem.
Our main contributions are as follows:
\begin{enumerate}
    \item We present the very first attempt to jointly model ASR and AAC tasks, and propose various E2E Transformer frameworks to solve this multi-task problem.
    \item Due to the lack of an audio dataset containing both transcript and caption labels, we prepare a synthetic multi-task dataset by combining clean speech samples and captioned non-speech samples.
    \item We carefully evaluate the proposed jointly trained models on varying levels of background sounds and compare them with the independently trained ASR and AAC models.
\end{enumerate}


\section{Independent Modeling}
\label{sec:ind-model}

\subsection{Automatic Speech Recognition}
\label{subsec:ASR}
\begin{figure}[t]
    \centering
    \includegraphics[clip, trim=0cm 10cm 3cm 0cm, width=\linewidth]{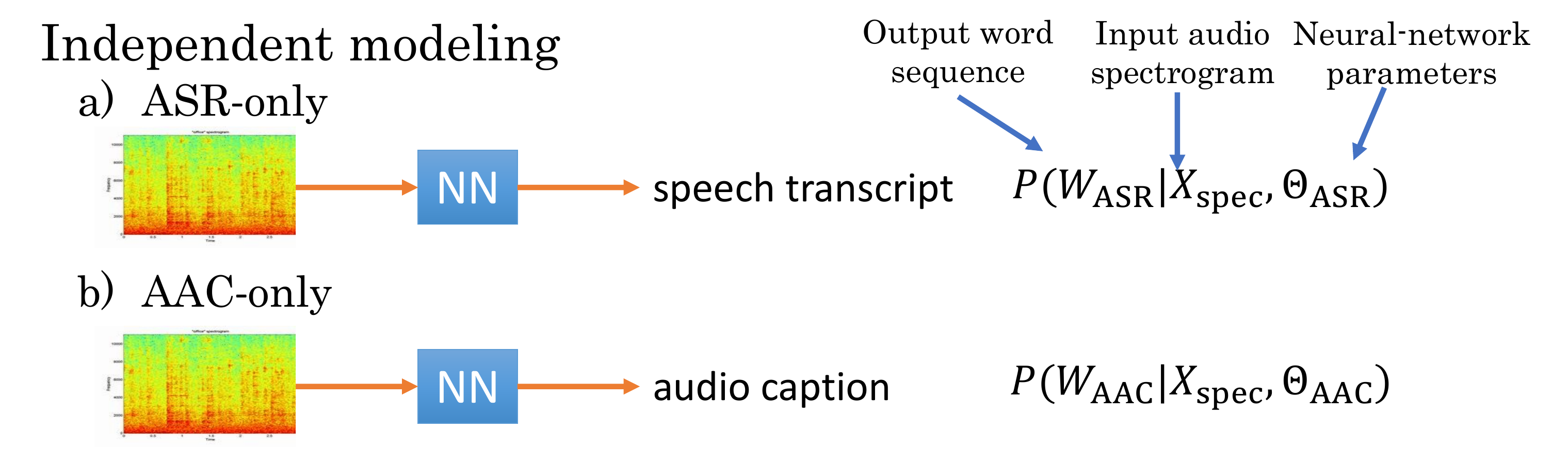}
    \caption{Typical independent modeling of E2E ASR and AAC tasks}
    \label{fig:indmodel}
\end{figure}

Popular E2E ASR systems are based on an encoder-decoder architecture, where the input acoustic features are decoded as speech transcripts.
Let $X_\mathrm{spec} \in \mathbb{R}^{F\times T}$ be an input spectrogram with frequency and time dimensions of $F$ and $T$ respectively, and $W_{\mathrm{ASR}}$ be the output word sequence. 
A neural network model with parameters $\Theta_\mathrm{ASR}$, aims to model the posterior distribution $P(W_{\mathrm{ASR}} | X_\mathrm{spec}, \Theta_\mathrm{ASR})$, as shown in Fig.~\ref{fig:indmodel}(a).
Some state-of-the-art models follow a Transformer encoder-decoder framework, optimized using both CTC loss and attention loss \cite{watanabe2018espnet}.
Typically, pretrained language models (LM) are integrated with the decoder during inference to form coherent word sequences \cite{hori2017advances,kannan2018analysis}.
However, as the purpose of this work is to compare the fundamental capability of independent and joint modeling approaches, we will not use LMs during inference.

\subsection{Automated Audio Captioning}
\label{subsec:AAC}
In contrast to ASR systems, which primarily focus on speech contents, AAC systems must model all the acoustic sources in an input signal and output an intelligible textual description (caption) \cite{kim2019audiocaps}.
Neural network models for the AAC task aim to model the posterior distribution $P(W_{\mathrm{AAC}} | X_\mathrm{spec}, \Theta_\mathrm{AAC})$, where $W_{\mathrm{AAC}}$ denotes the output word sequence in an audio caption, as shown in Fig.~\ref{fig:indmodel}(b).
As both ASR and AAC systems output word sequences, state-of-the-art AAC models also follow a Transformer based encoder-decoder framework, optimized using attention loss \cite{mei2021audio,narisetty2021_t6}.
Note that the CTC loss or RNN-Transducers are not particularly applicable to the AAC task, because the token sequence in a caption need not be temporally aligned with the input spectrogram frames.
In this work, the transcripts and captions are tokenized as a sequence of characters.


\section{Multi-task Dataset Synthesis}
\label{sec:dataset}
The inter-dependency of ASR and AAC tasks, in addition to their similar modeling techniques, motivates us to explore models that can be trained jointly on these two tasks. 
However an important challenge in realizing this joint modeling is the dearth of labeled real-world recordings with both a speech transcript and an audio caption.
We overcome this hurdle by synthetically mixing clean speech samples from the Wall Street Journal (WSJ) corpus \cite{paul1992design} with non-speech samples from AudioCaps dataset \cite{kim2019audiocaps}.
To support reproducible research, we will make our data generation scripts publicly available\footnote{\url{https://chintu619.github.io/Joint-ASR-AAC/}}.

WSJ corpus is a collective of the WSJ0 and WSJ1 datasets, totalling more than $37$k clean speech samples and their transcripts, uttered by roughly 300 speakers.
The AudioCaps dataset contains 46k audio samples paired with human-annotated captions, and is a subset of AudioSet (a large-scale collection of 10-second audio clips from YouTube) \cite{gemmeke2017audio}. 
Our goal is to prepare a dataset where an audio sample contains speech from a primary speaker in the presence of background noise.
Keeping with this goal, we identify and remove all the speech samples present in the AudioCaps dataset. 
This is done by filtering audio samples whose captions contain the sub-strings `speak' and `talk' (identifying words such as speaks, talking, etc.).
The filtering process discarded approximately 20k speech samples and retained $26$k non-speech audio samples.

We created a synthetic multi-task dataset by randomly mixing WSJ samples with non-speech AudioCaps samples. Given a speech signal $x_1$ and a non-speech signal $x_2$, we perform mixing as:
\begin{align}
    x_\mathrm{mix} = \hat{x}_1 + \gamma\cdot\hat{x}_2,\label{eq:asr-aac-mix}
\end{align}
where $\hat{(.)}$ represents an amplitude normalization to $[-1,1]$ range and $\gamma$ denotes a scalar mixing weight. 
The duration of audio samples from WSJ can be up to 25 seconds, while those from AudioCaps are roughly 10 seconds.
When $\hat{x}_1$ is longer than $\hat{x}_2$, we account for this duration mismatch during the mixing process in Eq.~(\ref{eq:asr-aac-mix}) by randomly choosing a segment of $\hat{x}_1$, such that the duration of chosen segment is equal to the duration of $\hat{x}_2$.

The transcript $W_{\mathrm{ASR}}$ of signal $x_1$ and caption $W_{\mathrm{AAC}}$ of signal $x_2$ are stored as labels for the normalized mixture signal $\hat{x}_\mathrm{mix}$. 
Although the addition of a clean speech event to the sounds present in $x_2$ modifies the textual description of $\hat{x}_\mathrm{mix}$, we overlook this issue from a practical standpoint. 
For instance, if $W_{\mathrm{AAC}}$ were to be `\textit{a dog is barking}', then a human annotated caption for $\hat{x}_\mathrm{mix}$ would probably be of the form `\textit{a person is speaking while a dog is barking}'.
Considering that all the mixture signals contain speech, their corresponding captions should ideally contain a phrase equivalent to `\textit{a person is speaking}'.
This inconsistency is common across all the mixture samples and can therefore be overlooked for our synthetic multi-task mixture dataset.

\section{Proposed Joint Modeling}
\label{sec:jointmodeling}
Given a set of noisy speech signals contaminated with various background sounds, we aim to train a single model capable of generating both the transcript and caption labels.
Inferring two output sequences from such a jointly trained model resembles an E2E multi-speaker ASR task \cite{chang2019end,kanda2020serialized}, without needing to accommodate for the permutation invariance among multiple speakers.

\begin{figure}[t]
    \centering
    \includegraphics[clip, trim=0cm 9.5cm 3cm 0cm, width=\linewidth]{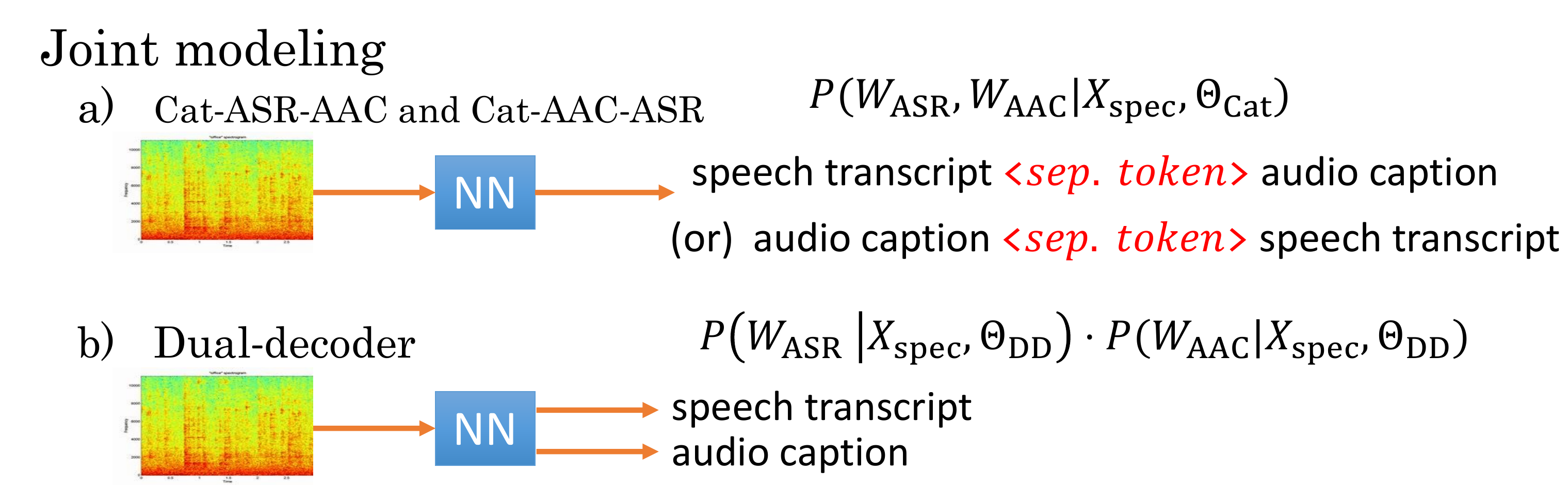}
    \caption{Joint modeling approaches of E2E ASR and AAC tasks}
    \label{fig:jointmodel}
\end{figure}

\subsection{Concatenating Output Sequences}
\label{subsec:jointconcat}
The independent ASR and AAC modeling frameworks discussed in Section~\ref{sec:ind-model} inherently follow a similar structure: acoustic feature encoding and word sequence decoding.
Exploiting this similarity and the serialized output training proposed in \cite{kanda2020serialized}, we propose an intuitive approach to our multi-task modeling: concatenating the word sequences $W_{\mathrm{ASR}}$ and $W_{\mathrm{AAC}}$ as follows:
\begin{align}
    W_\mathrm{Cat} &= \mathrm{Concat}(\; W_{\mathrm{ASR}}, \left<\textit{sep. token}\right>, W_{\mathrm{AAC}} \;) \label{eq:w-cat}\\
    W_\mathrm{Rev} &= \mathrm{Concat}(\; W_{\mathrm{AAC}}, \left<\textit{sep. token}\right>, W_{\mathrm{ASR}} \;) \label{eq:w-rev},
\end{align}
where $W_\mathrm{Cat}$ and $W_\mathrm{Rev}$ denote the concatenated word sequences, and $\left<\textit{sep. token}\right>$ denotes a separation token.
Eq.~(\ref{eq:w-cat}) signifies the scenario where $W_{\mathrm{ASR}}$ precedes $W_{\mathrm{AAC}}$ in the concatenated sequence, while Eq.~(\ref{eq:w-rev}) signifies a sequence concatenated in the reverse order.
For definitive detection of transition from $W_{\mathrm{ASR}}$ to $W_{\mathrm{AAC}}$ or vice-versa during model inference, we incorporate a special separation token $\left<\textit{sep. token}\right>$ during concatenation as shown in Fig.~\ref{fig:jointmodel}(a).

Given the input $X_\mathrm{spec}$ and model parameters $\Theta_\mathrm{Cat},\Theta_\mathrm{Rev}$ for $W_{\mathrm{Cat}}, W_{\mathrm{Rev}}$ respectively, we formulate the posterior distribution as: 
\begin{align}
    P(W_{\mathrm{Cat}} | X_\mathrm{spec},\Theta_\mathrm{Cat}) = \;& P(W_{\mathrm{ASR}} | X_\mathrm{spec},\Theta_\mathrm{Cat}) \cdot \nonumber\\ & P(W_{\mathrm{AAC}} | W_{\mathrm{ASR}},X_\mathrm{spec},\Theta_\mathrm{Cat}), \label{eq:joint-asr-aac}\\
    P(W_{\mathrm{Rev}} | X_\mathrm{spec},\Theta_\mathrm{Rev}) = \;& P(W_{\mathrm{AAC}} | X_\mathrm{spec},\Theta_\mathrm{Rev}) \cdot \nonumber\\ & P(W_{\mathrm{ASR}} | W_{\mathrm{AAC}},X_\mathrm{spec},\Theta_\mathrm{Rev}).\label{eq:joint-aac-asr}
\end{align}
We note that the formulations of Eq.~(\ref{eq:joint-asr-aac}) and Eq.~(\ref{eq:joint-aac-asr}) may appear as equivalent, but the evolving hidden state(s) of the neural network's recurrent architecture increases their dissimilarity with each decoded token.
Henceforth, we denote the frameworks modeling $W_{\mathrm{Cat}}$ and $W_{\mathrm{Rev}}$ as `Cat-ASR-AAC' and `Cat-AAC-ASR' respectively.

Although the proposed sequence concatenation approach aims to model the true joint distribution of $W_{\mathrm{ASR}}$ and $W_{\mathrm{AAC}}$, it suffers from several drawbacks.
The decoding process is likely to output sub-optimal solutions due to a significant increase in the length of output sequence.
Also, the constraint of generating a transcript and caption as part of the same output sequence removes the temporal alignment between $W_\mathrm{Cat}$ or $W_\mathrm{Rev}$ and $\hat{x}_\mathrm{mix}$.
In-turn we lose the ability of optimizing our model using the CTC loss.
Furthermore, the shared decoder is constrained to learn the unified token-transition probabilities within transcripts and captions, even though their individual token-transition probabilities need not be correlated.

\subsection{Dual Output Decoding}
\label{subsec:jointdualoutput}
Inspired by multi-task modeling in \cite{moritz2020all} and to alleviate the aforementioned drawbacks, we propose a shared encoder and dual-decoder framework to extract effective audio representations while separately decoding $W_{\mathrm{ASR}}$ and $W_{\mathrm{AAC}}$.
This approach assumes conditional independence between the transcript and caption given the input $X_\mathrm{spec}$ and dual-decoder model parameters $\Theta_\mathrm{DD}$ when modeling the joint probability distribution. Specifically, we have:
\begin{align}
    P(W_{\mathrm{ASR}},\;& W_{\mathrm{AAC}} | X_\mathrm{spec}, \Theta_\mathrm{DD}) = \nonumber\\
    & P(W_{\mathrm{ASR}}| X_\mathrm{spec}, \Theta_\mathrm{DD}) \cdot
    P(W_{\mathrm{AAC}}| X_\mathrm{spec}, \Theta_\mathrm{DD}).\label{eq:joint-dual-dec}
\end{align}
The flexibility of using two decoders allows the proposed model to individually learn the token transition distributions within transcripts and captions.
The exponentially expanded search space of concatenated output model is now restored to output sequences of regular length.
Finally, our proposed model also allows the use of the CTC loss when optimizing the ASR decoder as follows:
\begin{align}
    \mathcal{L} = \lambda \cdot \mathcal{L}_\mathrm{ctc} + (1-\lambda) \cdot \mathcal{L}_\mathrm{att},\label{eq:loss-ctc-att}
\end{align}
where $\mathcal{L}, \mathcal{L}_\mathrm{ctc}$ and $\mathcal{L}_\mathrm{att}$ denote the overall, CTC and attention losses respectively, and $\lambda$ denotes the interpolation factor \cite{watanabe2017hybrid}.

\section{Experiments}
\label{sec:experiments}
\subsection{Dataset Preparation}
\label{subsec:dataset-prep}
Following the synthetic multi-task dataset preparation procedure detailed in Section~\ref{sec:dataset}, we randomly mix the $37$k speech samples from WSJ with $26$k non-speech AudioCaps samples using mixing weight $\gamma$.
To mix the 836 total samples in development and evaluation splits of WSJ, we proportionately select 600 samples at random from AudioCaps that are not seen during training.
We choose 5 discrete values for $\gamma\in\{0.1, 0.2, 0.4, 0.6, 0.8\}$ to perform the above mixing process, aiming to evaluate our proposed methods at different levels of background sounds.
This process results in $37\text{k}\times 5\approx 187\text{k}$ training samples and roughly 4k dev and eval samples.

\subsection{Model Training and Evaluation}
\label{subsec:model-train-eval}
We trained the independent and proposed joint modeling frameworks using the ESPNet toolkit \cite{watanabe2018espnet}.
The models compared during our experiments are as follows:
\begin{enumerate}[leftmargin=*]
    \item ASR-only: trained to only output $W_{\mathrm{ASR}}$
    \item AAC-only: trained to only output $W_{\mathrm{AAC}}$
    \item Cat-ASR-AAC: outputs the word sequence $W_\mathrm{Cat}$ from Eq.~(\ref{eq:w-cat})
    \item Cat-AAC-ASR: outputs the word sequence $W_\mathrm{Cat}$ from Eq.~(\ref{eq:w-rev})
    \item Dual-decoder: trained to output both $W_{\mathrm{ASR}}$ and $W_{\mathrm{AAC}}$
\end{enumerate}
All the models discussed in this work follow the same network architecture to facilitate consistency during evaluation.
Our Transformer encoder has 12 layers with 2048 units each, and 4 attention heads of 256 dimension.
On the other hand, the Transformer decoder has 6 layers with 2048 units each.
The proposed dual-decoder model in Section~\ref{subsec:jointdualoutput} employs two decoders with the above specifications.
Our batch-size for training is set to 64, dropout to 0.1 and label smoothing weight to 0.1.
We use the Noam optimizer to train for 100 epochs and perform model parameter averaging over epochs with the best 10 validation scores.

Since CTC is not applicable to the concatenated output model, as discussed in Section~\ref{subsec:jointconcat}, we only perform experiments with the CTC loss for ASR-only and Dual-decoder models.
In this set of experiments, we set the interpolation factor $\lambda$ as 0.3 in Eq.~(\ref{eq:loss-ctc-att}). And during inference, we use beam-search with width 6 and CTC decoding weight of 0.3.
When training without CTC, we observed that inference with a beam width of 1 gave the best results.

All the trained models are evaluated over the dev and eval mixture samples for all the 5 levels of background sounds.
The generated speech transcripts are evaluated using traditional character and word error rates i.e., CER and WER.
The captions are evaluated using captioning metrics: CIDEr \cite{vedantam2015cider}, SPICE \cite{anderson2016spice} and SPIDEr \cite{liu2017improved}, where SPIDEr computes a simple average of CIDEr and SPICE scores.
The $n$-gram cosine similarity between generated and target captions is computed by CIDEr, while their semantic similarity after lemmatization is computed by SPICE.
Therefore, SPIDEr score balances the semantic and syntactic similarity during audio captioning.

\begin{table}[t]
\centering
\begin{tabular}{ l||c|c||c|c|c }
 \hline
 \textbf{Method} & CER & WER & CIDEr & SPICE & SPIDEr \\
 \hline\hline
 ASR-only & 11.4 & 24.0 & -- & -- & -- \\\hline
 AAC-only & -- & -- & 0.441 & 0.140 & 0.291 \\\hline
 Cat-ASR-AAC & 12.3 & 25.0 & 0.507 & 0.153 & 0.330 \\\hline
 Cat-AAC-ASR & \textbf{10.9} & \textbf{23.2} & \textbf{0.632} & \textbf{0.171} & \textbf{0.401} \\\hline
 Dual-decoder & 11.7 & 24.3 & 0.462 & 0.134 & 0.298 \\\hline
\end{tabular}
\caption{Comparison of speech and captioning metrics scores for all models trained without CTC and evaluated over the combined eval split of all mixing weights.}
\label{tab:results-overall}
\end{table}

\begin{table*}[t]
\centering
\begin{tabular}{ l||c|c|c|c|c||c|c|c|c|c }
 \hline
 \multirow{2}{6em}{\textbf{Method} (without CTC)} & \multicolumn{5}{c||}{\textbf{SPIDEr of dev-split}} & \multicolumn{5}{c}{\textbf{SPIDEr of eval-split}} \\
 \cline{2-11}
  & $\gamma=$ 0.1 & $\gamma=$ 0.2 & $\gamma=$ 0.4 & $\gamma=$ 0.6 & $\gamma=$ 0.8 & $\gamma=$ 0.1 & $\gamma=$ 0.2 & $\gamma=$ 0.4 & $\gamma=$ 0.6 & $\gamma=$ 0.8 \\
 \hline\hline
 AAC-only & 0.309 & 0.322 & 0.292 & 0.283 & 0.279 & 0.292 & 0.280 & 0.293 & 0.291 & 0.252 \\\hline
 Cat-ASR-AAC & 0.328 & 0.311 & 0.343 & 0.348 & 0.354 & 0.297 & 0.309 & 0.337 & 0.339 & 0.326 \\\hline
 Cat-AAC-ASR & \textbf{0.392} & \textbf{0.397} & \textbf{0.406} & \textbf{0.423} & \textbf{0.440} & \textbf{0.337} & \textbf{0.347} & \textbf{0.371} & \textbf{0.374} & \textbf{0.382} \\\hline
 Dual-decoder & 0.268 & 0.289 & 0.320 & 0.327 & 0.327 & 0.229 & 0.235 & 0.243 & 0.242 & 0.239 \\\hline
\end{tabular}
\begin{tabular}{ l||c|c|c|c|c||c|c|c|c|c }
 \hline
 \multirow{2}{6em}{\textbf{Method} (without CTC)} & \multicolumn{5}{c||}{\textbf{CER of dev-split}} & \multicolumn{5}{c}{\textbf{CER of eval-split}} \\
 \cline{2-11}
  & $\gamma=$ 0.1 & $\gamma=$ 0.2 & $\gamma=$ 0.4 & $\gamma=$ 0.6 & $\gamma=$ 0.8 & $\gamma=$ 0.1 & $\gamma=$ 0.2 & $\gamma=$ 0.4 & $\gamma=$ 0.6 & $\gamma=$ 0.8 \\
 \hline\hline
 ASR-only & \textbf{9.0} & 10.5 & 14.1 & \textbf{17.8} & \textbf{22.2} & 7.1 & 8.0 & 10.8 & 13.8 & \textbf{17.4} \\\hline
 Cat-ASR-AAC & 9.7 & 11.4 & 15.3 & 19.6 & 23.8 & 7.3 & 8.6 & 11.9 & 15.3 & 18.7 \\\hline
 Cat-AAC-ASR & 9.4 & \textbf{10.4} & \textbf{13.9} & 18.3 & 22.8 & \textbf{6.4} & \textbf{7.2} & \textbf{9.9} & \textbf{13.4} & 17.5 \\\hline
 Dual-decoder & 9.1 & 10.7 & 14.3 & 18.5 & 22.5 & 7.0 & 7.9 & 11.3 & 14.4 & 17.8 \\\hline
\end{tabular}
\begin{tabular}{ l||c|c|c|c|c||c|c|c|c|c }
 \hline
 \multirow{2}{6.07em}{\textbf{Method}\;\; (with CTC)} & \multicolumn{5}{c||}{\textbf{CER of dev-split}} & \multicolumn{5}{c}{\textbf{CER of eval-split}} \\
 \cline{2-11}
  & $\gamma=$ 0.1 & $\gamma=$ 0.2 & $\gamma=$ 0.4 & $\gamma=$ 0.6 & $\gamma=$ 0.8 & $\gamma=$ 0.1 & $\gamma=$ 0.2 & $\gamma=$ 0.4 & $\gamma=$ 0.6 & $\gamma=$ 0.8 \\
 \hline\hline
 ASR-only & 5.7 & 6.5 & 9.3 & 12.6 & \textbf{16.0} & \textbf{4.2} & 5.0 & 6.9 & \textbf{9.4} & \textbf{12.3} \\\hline
 Dual-decoder & \textbf{5.5} & \textbf{6.4} & \textbf{9.2} & \textbf{12.4} & 16.3 & \textbf{4.2} & \textbf{4.7} & \textbf{6.7} & 9.6 & 12.4 \\\hline
\end{tabular}
\caption{Comparison of models are trained without CTC (top, middle) and with CTC (bottom) and evaluated using CER (lower is better) and SPIDEr (higher is better) over the dev and eval splits for various mixing weights $\gamma\in\{0.1, 0.2, 0.4, 0.6, 0.8\}$.}
\label{tab:results-main}
\end{table*}

\section{Results}
\label{sec:results}
This section evaluates the performance of models trained independently for ASR, AAC tasks and compares them with our proposed jointly trained models.

\subsection{Independent vs. Joint Modeling}
\label{subsec:without_ctc}

The overall scores of speech and captioning metrics for all mixture weights $\gamma$ in Eq.~(\ref{eq:asr-aac-mix}), when training without the CTC is detailed in Table~\ref{tab:results-overall}.
Top and middle sections of Table~\ref{tab:results-main} summarizes the performance comparison when training without CTC and evaluating over all 5 mixing weights.
Expected trends include the increase of SPIDEr and CER scores with increasing mixture weight $\gamma$ of non-speech audio signal.
We also observe that CER for Cat-AAC-ASR model is better as compared to Cat-ASR-AAC.
The latter infers a speech transcript after generating the caption, thereby demonstrating the ability of Cat-AAC-ASR model to understand and utilize the background audio information when performing ASR.
Bottom section of Table~\ref{tab:results-main} shows significant improvements when using CTC for speech transcription.
It also empirically demonstrates the advantage of our joint modeling framework towards better speech recognition capability.

\subsection{Testing with Real-world Speech Recordings}
\label{subsec:realworld-testing}
We acknowledge that it would be premature to conclusively state the superiority of our multi-task modeling based solely on improvements seen with synthetic mixtures.
Therefore we additionally compare the model outputs over a few speech samples recorded in the presence of background sounds.
These recordings were chosen from the speech samples present in AudioCaps dataset, which were filtered-out during mixing process.

For an audio sample captured during a duck hunt gameplay\footnote{\url{https://youtu.be/8hSarhQXJbg?start=30\&end=40}}, the transcripts and captions generated from models trained without the CTC loss are compared with human annotations in Table~\ref{tab:results-realdata}.
The 10 second audio sample actually starts with some gun shots, followed by a person speaking and more gun shots, while ducks quack in the background.
Although the human annotated caption is short, our dual-decoder model is able to capture all the constituent audio sources.
On the other hand, all the ASR outputs seem mostly similar, with our proposed Cat-AAC-ASR and dual-decoder models slightly outperforming the ASR-only model.

\begin{table}[t]
\centering
\begin{tabular}{ p{6em}|p{18em} }
 \hline
 \textbf{Method} & \textbf{Generated Output(s)} \\
 \hline\hline
 ASR-only & \textcolor{cyan}{n. e. scale now that's a break job} \\\hline
 AAC-only & \textcolor{magenta}{a train approaches and blows a horn} \\\hline
 \multirow{2}{6.1em}{Cat-AAC-ASR} & \textcolor{cyan}{nice giel now that's a break job} \\
 & \textcolor{magenta}{a gun fires and a person whistles} \\\hline
 \multirow{2}{6em}{Dual-decoder} & \textcolor{cyan}{nise feel now that's a break job} \\
 & \textcolor{magenta}{gunshots fire and male voices with gunshots and blowing while a duck quacks in the background} \\\hline
 \multirow{2}{5.5em}{Human} & \textcolor{cyan}{nice kill, now that's a great shot} \\
  & \textcolor{magenta}{a man speaking and gunshots ringing out} \\\hline
\end{tabular}
\caption{Comparison of transcripts (blue) and captions (pink) between human annotations and generations from independently and jointly trained models for a speech sample recorded in real-world.}
\label{tab:results-realdata}
\end{table}

\section{Conclusions}
\label{sec:conclusions}
ASR and AAC tasks generate speech transcripts and descriptive audio captions respectively, but share a commonality of outputting coherent word sequences from a given audio signal.
As existing state-of-the-art models proposed for these tasks follow a similar Transformer based encoder-decoder framework, we explore the first known attempt at modeling ASR and AAC as a multi-task problem. 
Our proposed method takes advantage of the common Transformer frameworks and propose several joint modeling approaches.
As there are no currently known audio datasets which are labeled with both speech transcripts and audio captions, we prepare a multi-task synthetic dataset by mixing the clean speech samples with non-speech captioned audio samples.
Our extensive experiments over noisy speech samples with varying levels of background sounds, demonstrate the capability of our proposed multi-task models to match or even surpass the performance of independently trained ASR and AAC models.

\section{Acknowledgement}
\label{sec:ack}
This work was supported in part by Sony Group Corporation and used the Extreme Science and Engineering Discovery Environment (XSEDE) ~\cite{xsede}, which is supported by National Science Foundation grant number ACI-1548562. Specifically, it used the Bridges system ~\cite{nystrom2015bridges}, which is supported by NSF award number ACI-1445606, at the Pittsburgh Supercomputing Center (PSC).



\vfill
\pagebreak

\bibliographystyle{IEEEbib}
\bibliography{refs}

\begin{thebibliography}{10}

\bibitem{yu2016automatic}
Dong Yu and Li~Deng,
\newblock {\em Automatic Speech Recognition.},
\newblock Springer, 2016.

\bibitem{graves2014towards}
Alex Graves and Navdeep Jaitly,
\newblock ``Towards end-to-end speech recognition with recurrent neural
  networks,''
\newblock in {\em International conference on machine learning}. PMLR, 2014,
  pp. 1764--1772.

\bibitem{amodei2016deep}
Dario Amodei, Sundaram Ananthanarayanan, Rishita Anubhai, Jingliang Bai, Eric
  Battenberg, Carl Case, Jared Casper, Bryan Catanzaro, Qiang Cheng, Guoliang
  Chen, et~al.,
\newblock ``Deep speech 2: End-to-end speech recognition in english and
  mandarin,''
\newblock in {\em International conference on machine learning}. PMLR, 2016,
  pp. 173--182.

\bibitem{graves2012sequence}
Alex Graves,
\newblock ``Sequence transduction with recurrent neural networks,''
\newblock {\em arXiv preprint arXiv:1211.3711}, 2012.

\bibitem{li2020developing}
Jinyu Li, Rui Zhao, Zhong Meng, Yanqing Liu, Wenning Wei, Sarangarajan
  Parthasarathy, Vadim Mazalov, Zhenghao Wang, Lei He, Sheng Zhao, et~al.,
\newblock ``Developing {RNN-T} models surpassing high-performance hybrid models
  with customization capability,''
\newblock {\em arXiv preprint arXiv:2007.15188}, 2020.

\bibitem{chorowski2015attention}
Jan Chorowski, Dzmitry Bahdanau, Dmitriy Serdyuk, Kyunghyun Cho, and Yoshua
  Bengio,
\newblock ``Attention-based models for speech recognition,''
\newblock {\em arXiv preprint arXiv:1506.07503}, 2015.

\bibitem{chan2016listen}
William Chan, Navdeep Jaitly, Quoc Le, and Oriol Vinyals,
\newblock ``Listen, attend and spell: A neural network for large vocabulary
  conversational speech recognition,''
\newblock in {\em Proc. of ICASSP}. IEEE, 2016, pp. 4960--4964.

\bibitem{karita2019comparative}
Shigeki Karita, Nanxin Chen, Tomoki Hayashi, Takaaki Hori, Hirofumi Inaguma,
  Ziyan Jiang, Masao Someki, Nelson Enrique~Yalta Soplin, Ryuichi Yamamoto,
  Xiaofei Wang, et~al.,
\newblock ``A comparative study on {Transformer} vs {RNN} in speech
  applications,''
\newblock in {\em 2019 IEEE Automatic Speech Recognition and Understanding
  Workshop (ASRU)}. IEEE, 2019, pp. 449--456.

\bibitem{li2014overview}
Jinyu Li, Li~Deng, Yifan Gong, and Reinhold Haeb-Umbach,
\newblock ``An overview of noise-robust automatic speech recognition,''
\newblock {\em IEEE/ACM Transactions on Audio, Speech, and Language
  Processing}, vol. 22, no. 4, pp. 745--777, 2014.

\bibitem{haeb2019speech}
Reinhold Haeb-Umbach, Shinji Watanabe, Tomohiro Nakatani, Michiel Bacchiani,
  Bjorn Hoffmeister, Michael~L Seltzer, Heiga Zen, and Mehrez Souden,
\newblock ``Speech processing for digital home assistants: Combining signal
  processing with deep-learning techniques,''
\newblock {\em IEEE Signal processing magazine}, vol. 36, no. 6, pp. 111--124,
  2019.

\bibitem{moritz2020all}
Niko Moritz, Gordon Wichern, Takaaki Hori, and Jonathan Le~Roux,
\newblock ``All-in-one transformer: Unifying speech recognition, audio tagging,
  and event detection.,''
\newblock in {\em INTERSPEECH}, 2020, pp. 3112--3116.

\bibitem{kim2019audiocaps}
Chris~Dongjoo Kim, Byeongchang Kim, Hyunmin Lee, and Gunhee Kim,
\newblock ``{AudioCaps}: Generating captions for audios in the wild,''
\newblock in {\em Proc. of NAACL-HLT}, 2019, pp. 119--132.

\bibitem{drossos2017automated}
Konstantinos Drossos, Sharath Adavanne, and Tuomas Virtanen,
\newblock ``Automated audio captioning with recurrent neural networks,''
\newblock in {\em Proc. of WASPAA}. IEEE, 2017.

\bibitem{narisetty2021_t6}
Chaitanya Narisetty, Tomoki Hayashi, Ryunosuke Ishizaki, Shinji Watanabe, and
  Kazuya Takeda,
\newblock ``Leveraging state-of-the-art {ASR} techniques to audio captioning,''
\newblock Tech. {R}ep., DCASE2021 Challenge, July 2021.

\bibitem{mei2021audio}
Xinhao Mei, Xubo Liu, Qiushi Huang, Mark~D Plumbley, and Wenwu Wang,
\newblock ``Audio captioning transformer,''
\newblock {\em arXiv preprint arXiv:2107.09817}, 2021.

\bibitem{watanabe2018espnet}
Shinji Watanabe, Takaaki Hori, Shigeki Karita, Tomoki Hayashi, Jiro Nishitoba,
  Yuya Unno, Nelson Enrique~Yalta Soplin, Jahn Heymann, Matthew Wiesner, Nanxin
  Chen, et~al.,
\newblock ``Espnet: End-to-end speech processing toolkit,''
\newblock {\em arXiv preprint arXiv:1804.00015}, 2018.

\bibitem{hori2017advances}
Takaaki Hori, Shinji Watanabe, Yu~Zhang, and William Chan,
\newblock ``Advances in joint {CTC}-attention based end-to-end speech
  recognition with a deep {CNN} encoder and {RNN-LM},''
\newblock {\em arXiv preprint arXiv:1706.02737}, 2017.

\bibitem{kannan2018analysis}
Anjuli Kannan, Yonghui Wu, Patrick Nguyen, Tara~N Sainath, Zhijeng Chen, and
  Rohit Prabhavalkar,
\newblock ``An analysis of incorporating an external language model into a
  sequence-to-sequence model,''
\newblock in {\em Proc. of ICASSP}. IEEE, 2018, pp. 1--5828.

\bibitem{paul1992design}
Douglas~B Paul and Janet Baker,
\newblock ``The design for the wall street journal-based {CSR} corpus,''
\newblock in {\em Speech and Natural Language: Proceedings of a Workshop Held
  at Harriman, New York, February 23-26, 1992}, 1992.

\bibitem{gemmeke2017audio}
Jort~F Gemmeke, Daniel~PW Ellis, Dylan Freedman, Aren Jansen, Wade Lawrence,
  R~Channing Moore, Manoj Plakal, and Marvin Ritter,
\newblock ``Audio set: An ontology and human-labeled dataset for audio
  events,''
\newblock in {\em Proc. of ICASSP}. IEEE, 2017, pp. 776--780.

\bibitem{chang2019end}
Xuankai Chang, Yanmin Qian, Kai Yu, and Shinji Watanabe,
\newblock ``End-to-end monaural multi-speaker {ASR} system without
  pretraining,''
\newblock in {\em Proc. of ICASSP}. IEEE, 2019, pp. 6256--6260.

\bibitem{kanda2020serialized}
Naoyuki Kanda, Yashesh Gaur, Xiaofei Wang, Zhong Meng, and Takuya Yoshioka,
\newblock ``Serialized output training for end-to-end overlapped speech
  recognition,''
\newblock {\em arXiv preprint arXiv:2003.12687}, 2020.

\bibitem{watanabe2017hybrid}
Shinji Watanabe, Takaaki Hori, Suyoun Kim, John~R Hershey, and Tomoki Hayashi,
\newblock ``Hybrid {CTC}/attention architecture for end-to-end speech
  recognition,''
\newblock {\em IEEE Journal of Selected Topics in Signal Processing}, vol. 11,
  no. 8, 2017.

\bibitem{vedantam2015cider}
Ramakrishna Vedantam, C~Lawrence~Zitnick, and Devi Parikh,
\newblock ``Cider: Consensus-based image description evaluation,''
\newblock in {\em Proceedings of the IEEE conference on computer vision and
  pattern recognition}, 2015, pp. 4566--4575.

\bibitem{anderson2016spice}
Peter Anderson, Basura Fernando, Mark Johnson, and Stephen Gould,
\newblock ``Spice: Semantic propositional image caption evaluation,''
\newblock in {\em European conference on computer vision}. Springer, 2016, pp.
  382--398.

\bibitem{liu2017improved}
Siqi Liu, Zhenhai Zhu, Ning Ye, Sergio Guadarrama, and Kevin Murphy,
\newblock ``Improved image captioning via policy gradient optimization of
  spider,''
\newblock in {\em Proceedings of the IEEE international conference on computer
  vision}, 2017, pp. 873--881.

\bibitem{xsede}
J.~Towns, T.~Cockerill, M.~Dahan, I.~Foster, K.~Gaither, A.~Grimshaw,
  V.~Hazlewood, S.~Lathrop, D.~Lifka, G.~D. Peterson, R.~Roskies, J.~R. Scott,
  and N.~Wilkins-Diehr,
\newblock ``Xsede: Accelerating scientific discovery,''
\newblock {\em Computing in Science \& Engineering}, vol. 16, no. 5, pp.
  62--74, Sept.-Oct. 2014.

\bibitem{nystrom2015bridges}
Nicholas~A Nystrom, Michael~J Levine, Ralph~Z Roskies, and J~Ray Scott,
\newblock ``Bridges: a uniquely flexible {HPC} resource for new communities and
  data analytics,''
\newblock in {\em Proceedings of the 2015 XSEDE Conference: Scientific
  Advancements Enabled by Enhanced Cyberinfrastructure}, 2015, pp. 1--8.

\end{thebibliography}

\end{document}